# Ultrafast manipulation of mirror domain walls in a charge density wave

Alfred Zong[1†], Xiaozhe Shen[2†], Anshul Kogar[1], Linda Ye[1], Carolyn Marks[3], Debanjan Chowdhury[1], Timm Rohwer[1‡], Byron Freelon[1§], Stephen Weathersby[2], Renkai Li[2], Jie Yang[2], Joseph Checkelsky[1], Xijie Wang[2], Nuh Gedik[1*].

[1]Massachusetts Institute of Technology, Department of Physics, Cambridge, MA 02139, USA.
[2]SLAC National Accelerator Laboratory, Menlo Park, CA 94025, USA.
[3]Center for Nanoscale Systems, Harvard University, Cambridge, MA 02138, USA.
[†]These authors contributed equally to this work.
[‡]Present address: Center for Free-Electron Laser Science, DESY, Notkestraße 85, 22607 Hamburg, Germany.
[§]Present address: University of Louisville, Department of Physics and Astronomy, Louisville, KY 40292, USA.
[*]Corresponding author. Email: gedik@mit.edu.

**Abstract:** Domain walls (DWs) are singularities in an ordered medium that often host exotic phenomena such as charge ordering, insulator-metal transition, or superconductivity. The ability to locally write and erase DWs is highly desirable, as it allows one to design material functionality by patterning DWs in specific configurations. We demonstrate such capability at room temperature in a charge density wave (CDW), a macroscopic condensate of electrons and phonons, in ultrathin $1T$-TaS$_2$. A single femtosecond light pulse is shown to locally inject or remove mirror DWs in the CDW condensate, with probabilities tunable by pulse energy and temperature. Using time-resolved electron diffraction, we are able to simultaneously track anti-synchronized CDW amplitude oscillations from both the lattice and the condensate, where photo-injected DWs lead to a red-shifted frequency. Our demonstration of reversible DW manipulation may pave new ways for engineering correlated material systems with light.

## INTRODUCTION

Domains are ubiquitous in systems with long-range order and are identified by different configurations of the same order parameter. Intense studies of domain formation have not only underpinned important technologies such as nonvolatile memory (*1*), but also led to surprising discoveries at domain walls (DWs): metallicity in a ferroelectric or antiferromagnetic insulator (*2*, *3*), superconductivity in a ferroelastic oxide (*4*), and charge density waves (CDWs) in a 2D semiconductor (*5*). These unique properties at DWs arise because the order parameter is disrupted locally, leading to a renormalization of single-particle spectra and quasi-particle interactions. A particularly interesting ground state in which DWs occur is the CDW phase, where electrons and phonons cooperatively condense to form a superlattice below a transition temperature. Recent studies have unveiled novel electrical and optical properties at CDW DWs (*6–11*), which not only hold potential for CDW-based devices (*6*, *7*, *12*) but also yield insight into the fundamental question of how CDWs interact with other phases such as the Mott insulating state (*8*, *9*) and superconductivity (*10*, *13*).

Despite these recent advances, there has been limited progress in manipulating CDW DWs, with no reports of stable yet rewritable DWs at room temperature. To create (or remove) a CDW DW, the challenge is to provide sufficiently high energy to lodge (or dislodge) long-lived defects in the superlattice, without causing irreversible damage to the underlying crystal. Here, we used a single femtosecond pulse with a tailored fluence to locally inject or erase mirror DWs in a CDW, which were directly monitored by transmission electron diffraction (Fig. S1). The femtosecond pulse transiently perturbs the CDW and introduces topological defects (*14–20*), discontinuities that cannot be removed by smoothing the order parameter. These defects then follow a nonequilibrium pathway of relaxation into various DW structures. By carefully tuning the pulse fluence, we were able to eliminate these CDW DWs using a subsequent light pulse. The processes of DW injection and removal are entirely reversible.

In the present work, we study the octahedrally-coordinated polytype of the layered dichalcogenide, $1T$-$TaS_2$, because it has two degenerate CDW configurations, $α$ and $β$, which are related by mirror reflection. Fig. 1A illustrates the series of CDW transitions in resistivity and diffraction measurements. An incommensurate CDW (IC) first occurs below $T_{IC}$ = 550 K. As $1T$-$TaS_2$ is further cooled, a weakly first-order transition to a nearly-commensurate CDW (NC) occurs at $T_{IC-NC}$ = 354 K. In this transition, the previously incommensurate superlattice forms small patches of commensurate hexagrams with a $\sqrt{13}×\sqrt{13}$ periodicity (Fig. 1B) (*21*). In reciprocal space, the transition is marked by a rotation ($\phi$) of superlattice peaks away from the [100] direction (Fig. 1A, upper inset). Either clockwise ($α$) or counterclockwise ($β$) rotation can occur (Fig. 1A, lower inset), depending on the relative hexagram arrangements (Fig. 1B). Thus, the NC phase spontaneously breaks in-plane mirror symmetry, yielding the possibility of two equivalent domains. Cooling below $T_{NC-C}$ = 184 K results in a strongly first-order transition to a Mott-insulating commensurate CDW (C) phase, where hexagram clusters inherit their orientation from the NC phase. Despite the apparent equivalence between $α$ and $β$ configurations, we observed only one orientation for the entire sample in all twenty temperature cycles across $T_{IC-NC}$, consistent with previous reports (*22–24*). This observation can be attributed to an extrinsic lattice strain field that lifts the degeneracy between the two orientations (*25*).

## RESULTS
### Single-pulse manipulation
To locally inject $α/β$ DWs in a single-domain $1T$-$TaS_2$, we focused an 80-fs, 800-nm (1.55-eV) pulse onto a $150 \times 150$ μm$^2$ region of interest (ROI) in a 50-nm-thick sample, and probed the ROI by MeV transmission electron diffraction. Fig. 2A and 2C show the diffraction patterns before and after the exposure of a single pulse with an incident fluence of 7 mJ/cm$^2$ at room temperature. Two sets of satellite peaks appear in Fig. 2C, signaling the presence of $α/β$ DWs. The locality of DW injection was verified by probing a neighboring region unexposed to the pulse, where only $α$ satellites were visible as in Fig. 2A. These $α/β$ DWs are stable up to $T_{IC-NC}$ = 354 K and stay unchanged in atmosphere, as confirmed by a room temperature test for one week. There was no irreversible damage due to the laser pulse; a temperature cycle to the IC phase at 370 K and back to room temperature removed all $α/β$ DWs and restored a uniform $α$ domain as before (Fig. S2).

Remarkably, applying the same laser pulse to the $α/β$ state (Fig. 2C) erased the $β$ domains in the ROI, returning it to the original state (Fig. 2A, S5E). Such reversible switching between $α$ only and α/β states was repeated by more than 1,000 pulses on four different samples. The result was also robust with different combinations of pulse wavelength and duration (Fig. S2). As we discuss below, the switch is non-deterministic and its probability can be tuned by temperature and pulse fluence. Once switched, the sample remained in the α/β state upon cooling from the NC to the C phase (Fig. 2D). The transition temperature $T_{NC-C}$ is the same in both α and α/β states, again suggesting no pulse-inflicted sample degradation. A single pulse, up to the highest fluence attempted (11 mJ/cm$^2$), was unable to create or annihilate α/β DWs in the C phase.

The diffraction patterns in Fig. 2C and 2D do not discriminate between inter-layer and intra-layer DWs because of the large electron beam spot (90 × 90 μm$^2$, full-width at half maximum (FWHM)). To distinguish between the two possibilities, we performed selected area diffraction (SAD) on the same ROI of Fig. 2A–D using a 1.1-μm-diameter beam of a transmission electron microscope (TEM). We observed intra-layer DWs (Fig. 2E), with separate α (red circle) and β (blue circle) domains, divided by α/β DWs (yellow circle). Overlaid color masks denote the approximate domain locations. The yellow region suggests the presence of submicron domains, and we cannot rule out the possibility of inter-layer DWs in this area. The unmasked corner (Figure 2E, top right) shows the bright-field electron micrograph in grayscale, featuring bend contours due to underlying strain (*26*). We note the absence of correlation between locations of bend contours and α/β DWs, even under high magnification (Fig. S3). This suggests that pulse-induced α/β DWs are unrelated to any macroscopic lattice deformation.

**DW-dependent CDW amplitude oscillation**

To investigate how these DWs modify the underlying CDW order, we characterized its amplitude mode (AM) frequency in an α/β state. This mode arises from a broken translational symmetry during the CDW formation (*27*). It manifests as a breathing mode of the CDW hexagrams (Fig. 3C, inset). In a time-resolved diffraction experiment, the AM causes an oscillatory transfer of intensities between the central Bragg peaks and superlattice peaks of both domains (*25*, *28*), as confirmed in our measurement performed at 40 K (Fig. 3A, 3B, and S4). The intensity sum rule is obeyed through the exact π phase shift between the Bragg and superlattice peaks. We obtained the mode frequency after Fourier transforming the oscillatory parts (Fig. 3C and 3D), which is consistent with the AM frequency measured separately in the single-domain state under the same conditions (Fig. 3D, vertical line). In this case, α/β DWs did not modify the spectroscopic signature of the CDW.

By contrast, the AM significantly softened (Fig. 3E, S4B, S4D) in a repeated measurement: after warming the sample to room temperature, we switched the sample into the α state and then back into the α/β state; we then cooled down again to 40 K to re-measure the AM frequency. In the CDW phase, a soft AM is associated with a suppressed order parameter due to temperature (*29*), defects (*6*, *11*), or external perturbations (*18*). For example, significant discommensuration reduces the AM frequency from 2.25 THz in the C phase to 2.1 THz in the NC phase at the phase transition temperature $T_{C-NC}$ (*30*). Therefore, different frequencies in Fig. 3D and 3E are suggestive of two distinct concentrations and distributions of pulse-induced DWs (see schematics in insets). In Fig. 3D, very few α/β DWs are present, similar to regions masked in red and blue in Fig. 2E. In Fig. 3E, DWs are dense, similar to the yellow-masked region in Fig. 2E. This proposed DW configuration is further evidenced by a short CDW correlation length extracted from the peak width (Fig. S4E) (*25*). The significant presence of a DW network hence leads to a softer AM. A high-resolution microscopy measurement is required to confirm the exact distribution of DWs. Nonetheless, we stress that a single pulse can change the DW distribution, a key feature that enables domains to be reoriented.

**Defect-mediated nonthermal pathway for DW injection/removal**

To understand how a single pulse creates or annihilates $\alpha/\beta$ DWs, we examine the sequence of events upon photoexcitation. With sufficient pulse energy, the initial carrier excitation melts the NC phase within 1 ps (*17*). In the meantime, IC order nucleates in the discommensurate regions (*17*) and domains continue to grow over a nanosecond timescale (*15*, *16*). We verified the existence of this nonequilibrium IC phase by taking a snapshot of the diffraction pattern at a pump-probe delay of 1 ns (Fig. 4A, inset), when the IC phase has fully developed. Since the broken mirror symmetry is restored in the IC phase, we postulate that the transient transition to the IC phase is a necessary intermediate step for creating or removing DWs. The energy required for this ultrafast NC-to-IC transition thus represents the minimal energy barrier in any domain reorientation.

This energy barrier can be seen by examining the minimum fluence required for a non-vanishing probability of introducing $\alpha/\beta$ DWs, $P_+$ (*25*). In Fig. 4B, we plotted $P_+$ as a function of both fluence and temperature, and superimposed a guide to eye (mesh) by fitting the raw data (spheres) with a second-order two-dimensional polynomial. For each temperature, $P_+$ remains at zero below a threshold fluence (dashed line) because the pulse lacks the energy to excite the mirror-symmetric IC phase. The threshold fluence increases while temperature decreases, as the transient IC order is suppressed in favor of a residual NC order (Fig. 4A).

Beyond the minimum threshold, the probability $P_+$ increases with an increasing fluence (Fig. 4B). By contrast, the probability of removing $\alpha/\beta$ DWs, $P_-$, decreases with fluence (Fig. S5F). This growing tendency to create DWs at high fluence can be understood if we consider the following microscopic picture of DW creation and annihilation. Starting from a single-domain NC phase (Fig. 4C, left), a strong femtosecond pulse instantaneously suppresses the NC order while seeding many nucleation centers for the IC phase ($t \sim 1$ ps), where the local potential barriers of an NC-to-IC transition are among the lowest (*17*). Patches of IC order have distinct phases in the order parameter, and the energy deposited creates strong phase fluctuations that inhibit a complete coalescence of IC domains. Hence, a high concentration of topological defects remains when the IC order starts to relax ($t \sim 1–100$ ns). These CDW defects introduce local strains that break the $\alpha/\beta$ degeneracy at different locations, as sketched in the Landau-type potentials in Fig. 4C (*25*). Hence, as the system relaxes, both orientations develop in the NC phase (Fig. 4C, right) (*25*).

The erasure of DWs can be similarly instigated by another femtosecond pulse (Fig. 4C, bottom half), but with a reduced energy to limit the number of IC domains and pulse-induced defects. However, a minimum threshold fluence is still required to fully melt the NC order and reorient $\alpha$ and $\beta$ domains (e.g. $\sim 5$ mJ/cm$^2$ at 300 K) (Fig. 4A). In this case, the much-reduced defect concentration at the start of the IC-to-NC relaxation ($t \sim 1–100$ ns) favors a single NC domain, similar to a thermodynamic phase transition noted earlier. This proposed mechanism for domain switching is further supported by a simple model (*25*), where we relate the defect concentration in the transient IC phase to the size of reoriented domains. The simulated change in defect concentration after each pulse well captures the fluence dependences of both $P_+$ and $P_-$ (Fig. S5G).

**DISCUSSION**
In this picture, the key to manipulating mirror DWs lies in the control of the defect concentration in the transient IC phase. The mechanism differs from a conventional thermal quench scenario, where mirror domains may arise as sample quickly cools after pulse-induced heating (*25*). As the defect concentration depends on whether $\alpha/\beta$ DWs are present before the pulse arrival (*25*), the domain switching depends on sample history (Fig. S5E), an important characteristic in potential device applications. It is crucial to note that these are CDW and not lattice defects, so they can be effectively adjusted by optical pulses at various temperatures and fluences (Fig. 4B and S5F). In $1T$-TaS$_2$, these pulse-induced defects are useful for domain reorientation as they are long-lived ($\geq 1$ ns) and tunable by laser pulses, thanks to a rugged potential landscape with multiple local minima (*31*), where barriers are comparable to the pulse energy applied.

At these mirror DWs, the CDW amplitude and phase coherence are suppressed. In the Mott-insulating C phase, we therefore expect these DWs to host conducting channels in an insulating bulk (*3*, *8*, *9*), a promising property for nanoelectronics (*2*). To achieve custom functionality, one can pattern the light pulse, as done in holographic lithography, to create a specific network of DWs. Furthermore, the ability to switch between states with distinct orientations holds strong potential in storage applications such as a CDW-based memory (*32*). Here, $\alpha$ and $\beta$ domains act as "on" and "off" states, which are extremely stable at room temperature. More broadly, our results show that photoexcitation can be used as a reversible pathway between different states, providing a platform to realize unique properties in quantum matter through DW engineering.

**MATERIALS AND METHODS**
**Materials**
Single crystals of $1T$-TaS$_2$ were synthesized using the chemical vapor transport technique with I$_2$ as the transport agent following (*33*). Ta powder (Alfa Aesar, 99.97%) and S pieces (Sigma Aldrich, 99.998%) were first mixed with a molar ratio of 1:2.02, heated to 970 °C in an evacuated quartz tube, and subsequently quenched in cold water after two days. The resulting TaS$_2$ powder was sealed with I$_2$ again in an evacuated quartz tube and placed at the hot side of a temperature gradient from 920 °C to 820 °C in a three-zone furnace for the chemical vapor transport process. After two weeks, the quartz tube was quenched in water to avoid formation of the $2H$ phase. Thin, layered crystals with lateral size up to 1 cm could be found. Powder X-ray diffraction was performed to confirm that the obtained crystals were in the $1T$ phase.

An ultramicrotome fitted with a diamond blade was used to cleave $1T$-TaS$_2$ along the sulfur-sulfur plane (*34*), producing 50 nm-thick flakes of single crystals with an approximate dimension of $300 \times 300$ μm$^2$. Flakes were scooped from water onto standard TEM copper grids (300 lines/inch). The TEM grids were clamped to a copper holder, which was in good thermal contact with the cryostat cold head and the heater.

**Transport Measurement**
The resistivity of the $1T$-TaS$_2$ single crystal sample was measured with a Lakeshore 372 AC resistance bridge in a commercial cryostat using the four-probe method. The temperature ramping rate was kept as 3 K/minute for both cooling and warming. Constant voltage excitation was used to avoid heating at low temperature.

**Selected Area Diffraction (SAD)**
SAD was performed at room temperature using a commercial TEM (Tecnai $G^2$ Spirit TWIN, FEI) with 120 kV electron beam energy. An SAD aperture was inserted into the beam path, so only a 1.1 μm-diameter circular area on the sample was illuminated by the electrons. High resolution bright-field micrographs were taken with the same instrument, shown in Fig. 2E and S3.

**MeV Ultrafast Electron Diffraction (UED)**
The experiments on mirror DW manipulation and CDW amplitude mode characterization (Fig. 2–4, S4, and S5) were carried out in the MeV UED setup in the Accelerator Structure Test Area facility at SLAC National Laboratory (*35*). Fig. S1 shows a schematic of the setup. The 800-nm (1.55-eV), 80-fs pump pulse from a commercial Ti:sapphire regenerative amplifier (RA) laser (Vitara and Legend Elite HE, Coherent Inc.) were focused to a 150 × 150 μm$^2$ (FWHM) area in the sample at an incidence angle of ~ 5°. 3.1 MeV electron bunches with bunch length smaller than 150 fs were generated by radio-frequency photoinjectors. The electron beam was normally incident on the sample with a 90 × 90 μm$^2$ (FWHM) spot size. No sample degradation occurred due to the illumination of the MeV electron source, as confirmed by the same diffraction pattern and the same CDW transition temperature $T_{NC-C}$ measured over several days. The laser and electron pulses were spatially overlapped on the sample, and their relative arrival time was adjusted by a linear translation stage. The diffraction pattern was imaged by a phosphor screen (P-43) and recorded by an electron-multiplying charge-coupled device (EMCCD, Andor iXon Ultra 888). A circular through hole in the center of the phosphor screen allowed the passage of undiffracted electron beam to prevent camera saturation.

**keV UED**
The pulse-induced mirror DW creation and annihilation were reproduced in a separate UED setup (Fig. S2) at MIT with a different pulse wavelength and duration. The setup adopts a compact geometry (*36*). The 1038 nm (1.19 eV), 190 fs output of a Yb:KGW RA laser system (PHAROS SP-10-600-PP, Light Conversion) was focused to a 500 × 500 μm$^2$ (FWHM) area in the sample. The electron beam was generated by focusing the fourth harmonic (260 nm, 4.78 eV) to a gold-coated sapphire in high vacuum ($< 4 \times 10^{-9}$ torr). Photoelectrons excited were accelerated to 26 kV in a dc field and focused to an aluminum-coated phosphor screen (P-46) by a magnetic lens, with a 270 × 270 μm$^2$ (FWHM) beam spot at the sample position. Diffraction patterns were recorded by a commercial intensified CCD (iCCD PI-MAX II, Princeton Instruments). Data presented in Fig. 1A (inset) and Fig. S2 were taken with this setup.

**Acknowledgments:** We thank E. Baldini for insightful discussions. A.Z. and A.K. thank Y. Zhang for assisting the SAD measurement at the MRSEC Shared Experimental Facilities at MIT, supported by the National Science Foundation under award number DMR-14-19807. N.G., A.Z., A.K., T.R., and B.F. acknowledge support from the U.S. Department of Energy, BES DMSE (keV UED at MIT) and from the Gordon and Betty Moore Foundation's EPiQS Initiative, Grant GBMF4540 (data analysis and manuscript writing); X.W., X.S., S.W., R.L. and J.Y. acknowledge support from the U.S. Department of Energy BES SUF Division Accelerator & Detector R&D program, the LCLS Facility, and SLAC under contract No.'s DE-AC02-05-CH11231 and DE-AC02-76SF00515 (MeV UED at SLAC); J.C. and L.Y. acknowledge support from the Gordon and Betty Moore Foundation EPiQS Initiative, Grant GBMF3848 (sample growth and characterization). D.C. acknowledges the postdoctoral fellowship support from the Gordon and Betty Moore Foundation, under the EPiQS initiative, Grant GBMF4303, at MIT (data interpretation).

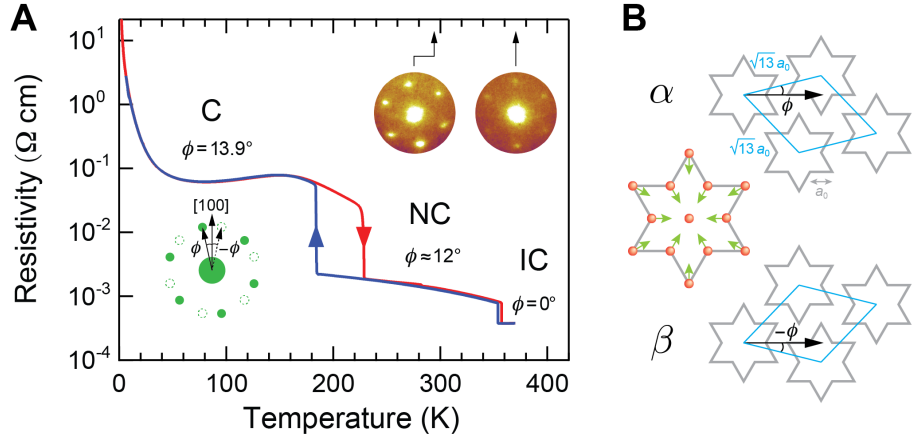

**Figure 1. Structural phases of 1$T$-TaS$_2$ and its two mirror-symmetric commensurate states.** (**A**) Temperature-dependent resistivity measured during cool-down (blue) and warm-up (red), showing hysteretic discontinuities as 1$T$-TaS$_2$ transitions through incommensurate (IC), nearly-commensurate (NC), and commensurate (C) charge density wave (CDW) phases. The triclinic phase during warmup (223 K to 283 K) is omitted. Lower inset, schematic diffraction pattern in CDW phases, with a central Bragg peak surrounded by six first-order superlattice peaks. Filled (dashed) circles represent $\alpha$ ($\beta$) orientation in the C or NC phase. Upper inset, static diffraction patterns of the (2 0 0) Bragg and satellite peaks taken at 295 K and 370 K. Only the $\alpha$ domain is present throughout the sample. (**B**) Schematics of in-plane atomic arrangements in commensurate regions for $\alpha$ (top) or $\beta$ (bottom) state that breaks the in-plane mirror symmetry. Orange spheres denote Ta atoms, which form clusters of regular hexagrams. Blue diamonds represent unit cells of the $\sqrt{13}\times\sqrt{13}$ superlattice.

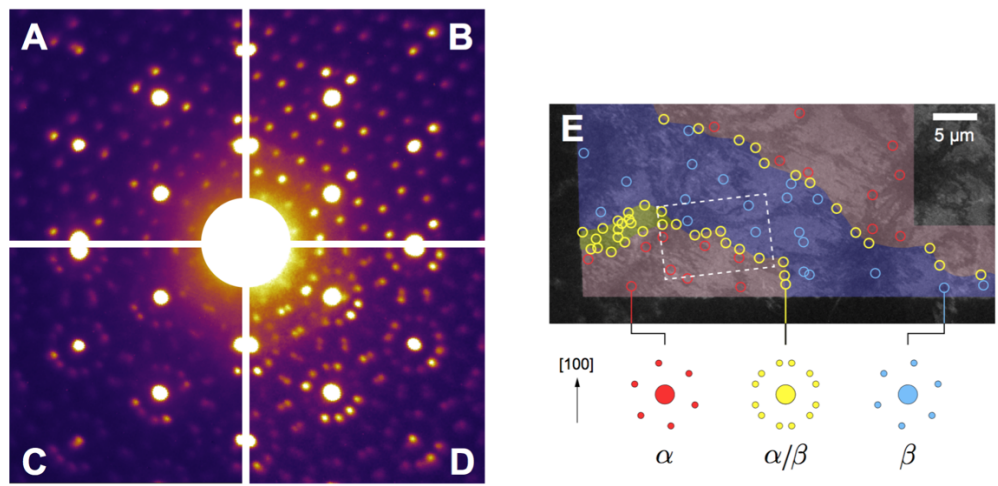

**Figure 2. Diffraction and microscopy of switchable domains by a single femtosecond pulse.** (**A** and **C**) $\alpha$ and $\alpha/\beta$ states of the NC phase at 300 K, respectively, which can be repeatedly switched from one to the other after applying a single 800-nm (1.55-eV), 80-fs pulse at 7 mJ/cm$^2$ incident fluence. (**B** and **D**) the corresponding $\alpha$ and $\alpha/\beta$ states in the C phase at 40 K, which can no longer be switched into the other by a single pulse up to the highest fluence attempted (11 mJ/cm$^2$). (**E**) Room-temperature, bright-field transmission electron micrograph taken in the pulse-induced $\alpha/\beta$ state. Selected area diffractions (SADs) were performed at each circle, whose diameter indicates a 1.1 μm beam size. Schematic SAD patterns are shown at the bottom. Overlaid red and blue masks indicate approximate locations of $\alpha$ and $\beta$ domains. The yellow mask (bottom left) suggests a region with submicron domains. The top right corner is not masked to expose the grayscale micrograph where bend contours are clearly visible. Fig. S3 shows a high magnification micrograph of the dashed rectangle. Black edges at the bottom and left are copper frames of the supporting TEM grid.

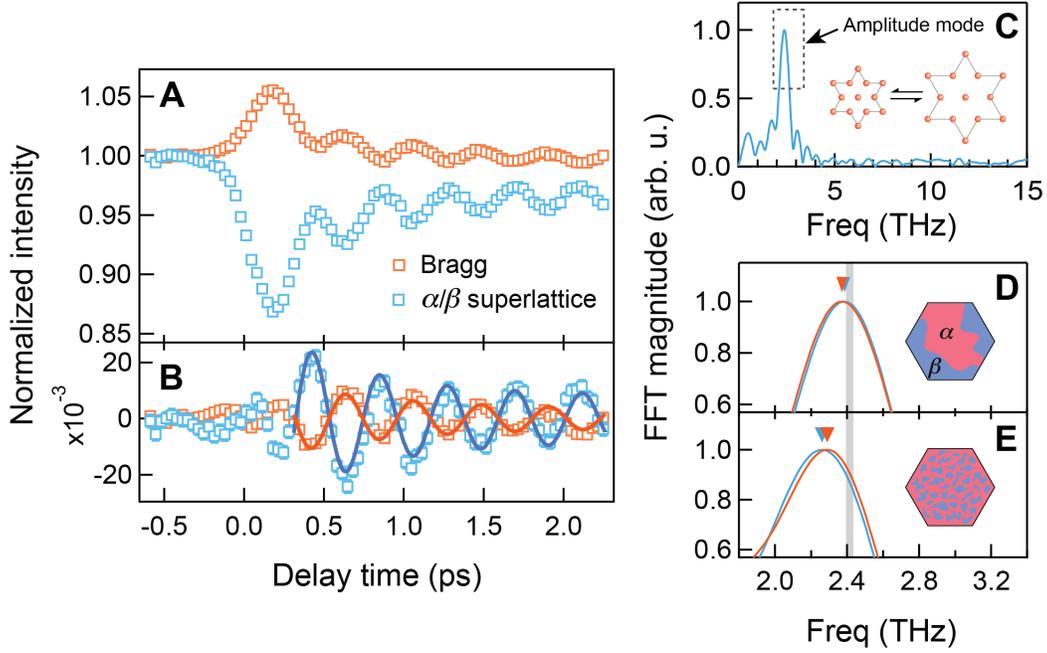

**Figure 3. Coherent excitation of the amplitude mode (AM) in *α/β* states of the C phase.**
(**A**) Integrated intensities of Bragg and superlattice peaks of both domains as a function of pump-probe delay time. Intensities are normalized by values before pump incidence. (**B**) The coherent, oscillatory part of the intensity after subtracting a fitted single-exponential, incoherent part from (**A**). Solid curves are fits to an exponentially decaying cosine function. (**C**) The Fourier transformed spectrum of the superlattice intensity in (**B**), featuring the prominent AM peak. The FFT was computed with zero padding and normalized to the peak value. Uncertainties reflect fitting errors to an exponentially decaying cosine. Inset, a schematic of the AM, corresponding to a breathing mode of hexagrams. (**D**) A zoomed-in view of the dashed rectangle in (**C**). Color codes are the same as used in (**A**). Peak positions are marked by inverted triangles. The gray vertical line indicates the AM frequency of a single-domain sample measured separately under the same condition. Widths of the triangles and the vertical line indicate frequency uncertainties. (**E**) The same as in (**D**), but after switching the *α/β* state to a single domain and back to an *α/β* state at room temperature. The AM markedly softened. Insets in (**D**) and (**E**), schematic DW distributions, where DWs are sparse in (**D**) but dense in (**E**). All data were taken at a base temperature of 40 K and an incident fluence of 1 mJ/cm$^2$.

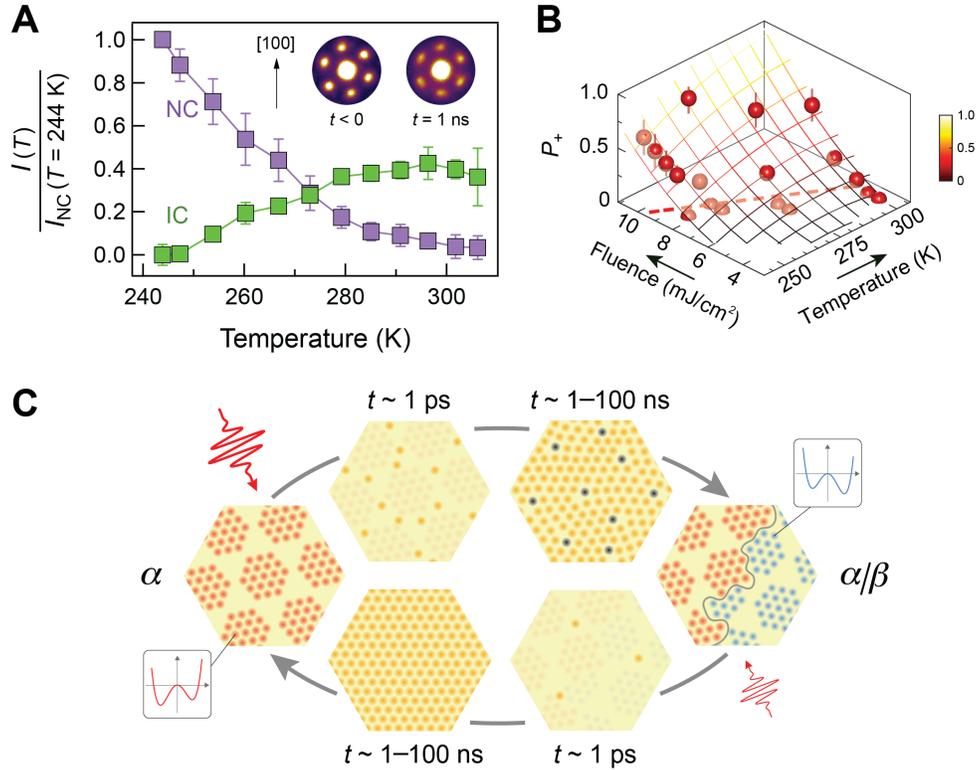

**Figure 4. Microscopic mechanism and probability of domain switching.** (**A**) Normalized integrated intensities of NC and IC peaks at various temperatures, taken at a pump-probe delay of 1 ns and an incident fluence of 5 mJ/cm². Inset, room temperature diffractions taken before photoexcitation ($t < 0$) and at 1 ns delay, showing the nonequilibrium NC-to-IC transition. Data were averaged over (3 0 0) and its five symmetry equivalent peaks. (**B**) Fluence- and temperature-dependence of $P_+$, the probability of injecting DWs by switching from an $\alpha$ only to an $\alpha/\beta$ state with a single pulse. Arrows indicate directions of increasing temperature and fluence. The mesh is a guide to eye, obtained by fitting the raw data (spheres) to a second-order two-dimensional polynomial. Dashed line marks the minimum threshold fluence for a nonzero $P_+$. (**C**) A real-space schematic of creating (upper half) and annihilating (lower half) mirror DWs. Periodic patterns represent superlattices in the NC (red, $\alpha$; blue, $\beta$) and the IC phase (yellow). Black dots mark the positions of unannihilated topological defects in the transient IC phase before it relaxes to the equilibrium NC phase. Landau-type potentials indicate the breaking of local $\alpha/\beta$ degeneracy in different parts of the sample (*25*). See text for a description of the event sequence.